\documentclass[twocolumn]{jpsj2}

\def\JA{J_{\rm A}}
\def\JF{J_{\rm F}}

\def\e{\rm e}
\def\c{\rm c}

\def\const{\rm const}
\def\mst{m_{\rm st}}

\def\simgeq{\mbox{\raisebox{-1.0ex}{$\stackrel{>}{\sim}$}}}
\def\ln{\mbox{ln}}
\def\v#1{\mbox{\boldmath$#1$}}
\def\clcl{${\mbox{Cu}}<^{\mbox{Cl}}_{\mbox{Cl}}>{\mbox{Cu}}$ }
\def\brbr{${\mbox{Cu}}<^{\mbox{Br}}_{\mbox{Br}}>{\mbox{Cu}}$ }
\def\clbr{${\mbox{Cu}}<^{\mbox{Br}}_{\mbox{Cl}}>{\mbox{Cu}}$ }
\def\runtitle{Random Magnetism in $S=1/2$ Heisenberg Chains with Bond
Alternation and Randomness on the Strong Bonds}
\def\runauthor{Kazuo \textsc{Hida}}

\title{%
Random Magnetism in $S=1/2$ Heisenberg Chains with Bond Alternation and 
Randomness on the Strong Bonds}

\author{%
Kazuo \textsc{Hida}\thanks{E-mail: hida@phy.saitama-u.ac.jp}
}

\inst{%
Department~of~Physics,
 Faculty of Science,\\ Saitama University, Saitama, Saitama 338-8570
}

\recdate{\today}

\abst{%
The $S=1/2$ Heisenberg chains  with bond alternation and randomness on the
strong bonds are studied by the density matrix renormalization group method.
It is assumed that the odd-th bond is antiferromagnetic with strength $J$
and even-th bond can take the values $\JA$ and $\JF$ $ (\JA > J > 0 > \JF)$
randomly. The ground state of this model interpolates between the Haldane
and dimer phases via a randomness dominated intermediate phase.  Based on
the scaling of the low energy spectrum and mean field treatment of the
interchain coupling, it is found that the  magnetic long range order is
induced by randomness in the intermediate regime. In the magnetization
curves, there appears a plateau at the fractional value of the saturated
magnetization. The fine structures of the magnetization curves and low
energy spectrum are understood based on the cluster picture. The relation
with the recent experiment for (CH$_3)_2$CHNH$_3$Cu(Cl$_x$Br$_{1-x})_3$ is
discussed.
}

\kword{%
random quantum spin chain, bond alternation, DMRG, disorder induced order,
magnetization plateau
}
\begin{document}

\maketitle

\section{Introduction}
In the recent studies of quantum many body problem, the one-dimensional
random quantum spin systems have been attracting a renewed interest from
theoretical and experimental
viewpoints.\cite{manaka,manakanew,uchi,uchiprog,fuku,azuma,uchi2,yasu1,df1,k
h1,hidap,cabra,totsuka,vilar,naka} 

Among them, the phenomenon of the disorder induced order have been widely
investigated. Experimentally, Uchinokura and coworkers\cite{uchi,uchiprog}
have found the antiferromagnetic ordered phase in  Zn, Mg, Si-doped
spin-Peierls compound CuGeO$_4$ as one of the earliest examples of this type
of phenomenon.  The theoretical explanation is given by Fukuyama and
coworkers\cite{fuku} using the bosonization approach. 

Similar phenomena are observed in Zn-doped SrCu$_2$O$_3$\cite{azuma} which
is the quasi-1-dimensional $S=1/2$ ladder system and Mg-doped
PbNi$_2$V$_2$O$_8$\cite{uchiprog,uchi2} which is the quasi-1-dimensional
$S=1$ Haldane gap system. Correspondingly, the effect of the bond and site
randomness on the spin gapped quasi-1-dimensional $S=1/2$ and $S=1$
antiferromagnets are studied by the quantum Monte Carlo method by Yasuda and
coworkers\cite{yasu1}. They have also found the randomness induced long
range order for appropriate range of randomness.

Recently, Manaka and coworkers\cite{manaka,manakanew} studied the magnetic
and thermal properties of the quasi-one-dimensional compound
(CH$_3)_2$CHNH$_3$Cu(Cl$_x$Br$_{1-x})_3$ (IPACu(Cl$_x$Br$_{1-x}$)). For
$x=0$, this material is the $S=1/2$ antiferromagnetic-antiferromagnetic
alternating Heisenberg chain (AF-AF chain) whose ground state is the  dimer
phase\cite{manakaaf} while for $x=1$ it is  the $S=1/2$
ferromagnetic-dominant ferromagnetic-antiferromagnetic alternating
Heisenberg chain (F-AF chain) with Haldane ground
state\cite{manakaf,manakasat}. In the intermediate concentration regime,
however, both the Haldane and dimer phases are destroyed by randomness.
Remarkably, Manaka and coworkers\cite{manaka} experimentally found the
magnetically ordered phase for $0.44 < x <0.87$. It should be also remarked
that the energy gap estimated from the temperature dependence of the
specific heat and susceptibilty remains finite even in the close
neighbourhood of the critical concentration. 

Modivated by this experiment, we investigate the random $S=1/2$ Heisenberg
chain whose Hamiltonian is given by,
\begin{equation}
\label{ham1}
H = \sum_{i=1}^{N}J  \v{S}_{2i-1} \v{S}_{2i} + \sum_{i=1}^{N}J_i  \v{S}_{2i}
\v{S}_{2i+1},
\end{equation}
where $J > 0$ and $J_i=\JF (< 0)$ with probability $1-p$ and  $J_i=\JA (>0)$
with probability $p$.
The ground state of this model interpolates between the Haldane phase
($p=0$) and dimer phase ($p=1$).

As discussed by Manaka and coworkers\cite{manakanew}, the relation between
$x$ and $p$ is not trivial because the exchange paths between two Cu ions are
bibridged bonds. For $x=1$, the ferromagnetic bonds are \clcl bonds, and for
$x=0$, the strong antiferromagnetic bonds are \brbr bonds. The addition of
the Br ions into the $x=1$ chain induces the \clbr bonds which are absent in
the pure chains. We assume, however, that the \clbr bonds are strongly
antiferromagnetic because otherwise the Haldane phase cannot be destroyed
and no magnetic order can take place  for relatively small Br concentration
($1-x \simeq 0.13$) as observed in the experiment. In this case, the
concentration of the ferromagnetic bond $1-p$ is equal to $x^2$. Further, we
may safely ignore the \brbr bonds in the concentration regime  $1-x << 1$ which we are mainly concerned in this paper. Actually at the critical
concentration $x=0.87$, the concentration of \brbr bonds is $(1-x)^2 \sim
0.017$ which is sufficiently small. We can therefore set $p = 1-x^2 $ in
this regime. In the following, we take $\JF= -2J$ following the experimental
estimation\cite{manaka} and $J = 1$ to fix the energy unit. 

In the present system, the spin gap state in the pure system is intrinsic.
Therefore the mechanism of the disorder induced order is not related with
the lattice degrees of freedom as in the spin-Peierls
systems\cite{uchi,uchiprog,fuku}. The situation is somewhat similar to the
site depleted 2-dimensional $S=1$  antiferromagnet studied by Yasuda and
coworkers\cite{yasu1} if $\JF$ and $\JA$ bonds are infinitely strong.
However, as explained later, the finiteness of these bonds introduces
interesting fine structures in the single chain properties which should be
observable in experiments.
\begin{figure}
\centerline{\includegraphics[width=70mm]{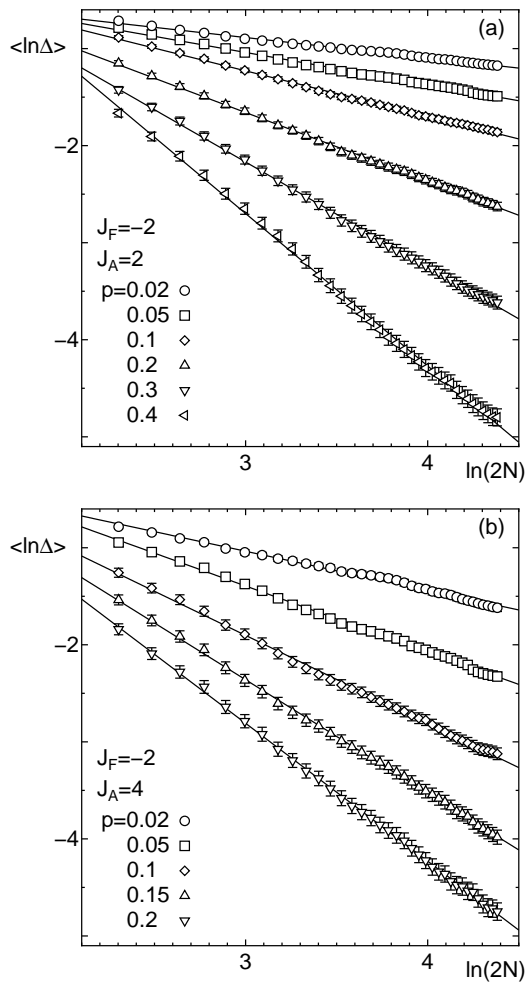}}
\caption{The system size-dependence of $<\ln\Delta>$ for (a) $\JA=2$ and (b)
$\JA=4$ with $\JF=-2$.  The solid line is the fit to the relation $\ln
\Delta = \const. - z \ln N$.}
\label{ndep}
\end{figure}

This paper is organized as follows. In the next section, the single chain
properties of the present model are discussed. In  subsection 2.1, the
low energy spectrum of the single chain is calculated and shown to have
quantum Griffiths singularity. The physical origin of this behavior is also
explained in the cluster picture. In subsection 2.2, the magnetization curve of this model is
presented and the randomness induced plateau is shown to appear. The fine structures of the magnetization curve are also explained in terms of the cluster picture. In the third section, the effect of interchain coupling is studied by means of
the interchain mean field approximation. It is shown that the disorder
induced magnetic order takes place in appropriate concentration regime. The
final section is devoted to summary and discussion. The brief report on the main results of this work is published in ref. \citen{hidap}.

\section{Single Chain Properties}

\subsection{Low Energy Excitation Spectrum}

\begin{figure}
\centerline{\includegraphics[width=70mm]{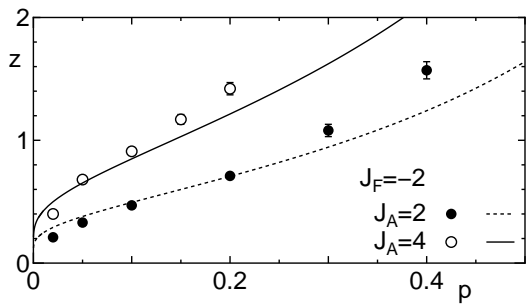}}
\caption{The  $p$-dependence of the dynamical exponent $z$. The circles are
the estimation from the system size dependence of $<\ln \Delta >$. The lines
are $z=\alpha/\mid\ln p\mid$ where $\alpha$ is determined from the $q$
dependence of $\Delta_q$.}
\label{exp}
\end{figure}

To support the long range order in the coupled chain systems, enough number
of low energy states are required in the spectrum of the constituent single
chains. We therefore calculate the size dependence of the energy gap $\Delta$
of the random ensemble of spin chains described by Eq. (\ref{ham1}). Due to
computational reason, we concentrate on the regime of small $p$ (small
$1-x$) in the following. Figure \ref{ndep} shows the system size dependence
of the average of the logarithm of the energy gaps $\Delta$ with $\JF=-2$
and (a) $\JA=2$ and (b) 4 for $10 \leq 2N \leq 80$ calculated by the density
matrix renormalization group (DMRG) method\cite{kh1,wh1}. The average is
taken over 400 samples. In this parameter range, these curves are fitted
well by $<\ln \Delta> \sim \mbox{const.}-z\ln N$ where $z$ defines the
dynamical exponent. This implies that the energy gap is scaled by $N^{-z}$
which is a typical behavior of the quantum Griffiths phase\cite{df1,kh1}. If
we assume such scaling holds for all low energy  excitation spectrum, the
distribution function of the low lying excitation energies scales as 
\begin{equation}
P(\Delta)d\Delta = N^z f(N^z\Delta)d\Delta
\end{equation}
where $P(\Delta)d\Delta$ is the number of states with excitation energy in
the range $[\Delta, \Delta+d\Delta]$. For large $N$, $P(\Delta)$ should be
proportional to the system size $N$, so that $f(x) \rightarrow x^{1/z-1}$ as
$x \rightarrow \infty$. Thus, in the thermodynamic limit, we have 

\begin{equation}
P(\Delta)d\Delta \propto N \Delta^{1/z-1}d\Delta.
\label{grif}
\end{equation}
Therefore the number of the low energy states diverges if $z > 1$. As shown
in Fig. \ref{exp}, $z$ increases with $p$ and becomes larger than unity
above a critical value $p_{\c}$ where $p_c \simeq 0.3$ for  $\JA=2$ and
$p_c \simeq 0.1$ for $\JA=4$. We can expect that the long range order would
be stabilized for $p > p_{\c}$ if the interchain coupling is switched on. 

\begin{figure}
\centerline{\includegraphics[width=70mm]{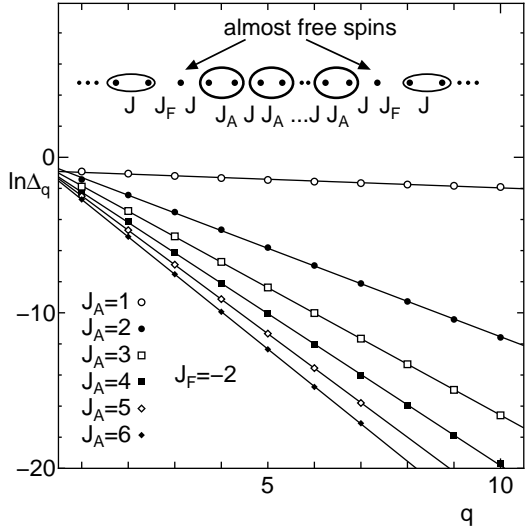}}
\caption{The  $q$-dependence of $\Delta_q$. The inset shows the $q$-cluster.}
\label{dlteff}
\end{figure}

These features of the  low energy spectrum can be understood in the
following way. Let us consider a cluster consisting of $q$ $\JA$-bonds and
$q-1$ $J$-bonds embedded in the F-AF chain as depicted in the inset of Fig.
\ref{dlteff}. This is called  '$q$-cluster' in the following. The middle
$2q$ spins form a tightly bound  singlet cluster. The two spins connected to
both ends of this cluster by $J$-bonds are almost free but weakly coupled
with each other mediated by the quantum fluctuation within the strongly
coupled cluster.  For small $p$, we can regard the whole system as a random
assembly of $q$-clusters. The $q$-dependence of the singlet triplet gaps
$\Delta_q$ of $q$-clusters calculated by the DMRG method is well fitted by
$\Delta_q \simeq \Delta_0 \e^{-\alpha q}$ as shown in Fig. \ref{dlteff}.  On
the other hand, the number of $q$-clusters in a chain  is proportional to
$Np^q(1-p)^2$. Eliminating $q$, the number $P(\Delta)d\Delta$ of the
$q$-clusters with energy gap between $\Delta$ and $\Delta + d\Delta$ behaves
as Eq. (\ref{grif})  with 
\begin{equation}
z= \alpha/\mid\ln p \mid.
\label{zvalue}
\end{equation}
This value of $z$ is plotted by the solid and dotted lines in Fig.
\ref{exp}. The agreement with the values obtained by  fitting the numerical
data of energy gap in Fig. \ref{ndep} is good for small $p$. 
As $p$ becomes larger, the interference between the clusters would prevent
the above simple-minded interpretation.

\begin{figure}
\centerline{\includegraphics[width=70mm]{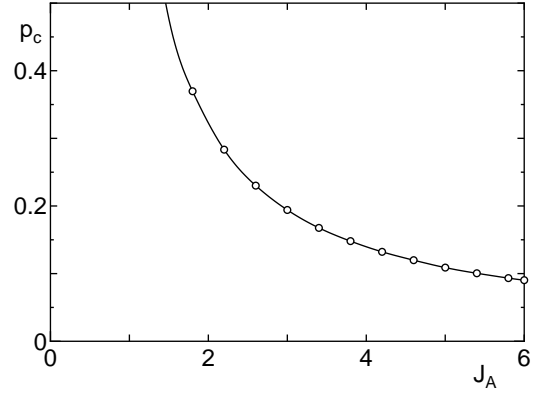}}
\caption{The parameter regime in which the low energy spectrum has divergent
singularity.}
\label{pc}
\end{figure}

Using the formula (\ref{zvalue}), we can estimate the parameter regime in
which the low energy spectrum has divergent singularity as  $p > p_c \equiv
\exp(-\alpha)$. The critical value  $p_c$ is plotted against $\JA$ in Fig.
\ref{pc} using the values of $\alpha$ obtained from Fig. \ref{dlteff}. For
$p > p_c$, it is possible that the long range order is stabilized in the
presence of interchain coupling. 

The above picture explicitly demonstrates that the low energy excitation of
the present model is dominated by the large size clusters and the
characteristic size of the clusters increases by the power law with the
decrease of the energy scale. This is a typical feature of the Griffiths
phase. This singular excitation spectrum is reflected in the low temperature
magnetic susceptibility $\chi$ and magnetic specific heat $C$ as  $\chi \sim
T^{1/z-1}$ and $C \sim T^{1/z}$.

\begin{figure}
\centerline{\includegraphics[width=70mm]{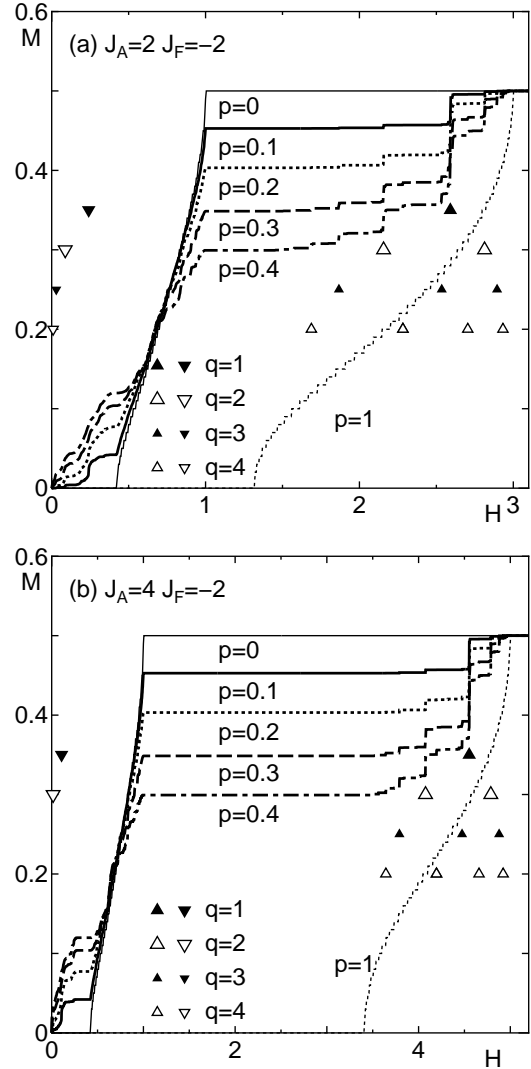}}
\caption{Magnetization curves for (a) $\JA=4$ and (b)  $\JA=2$ with
$\JF=-2$  and $2N=100$ for various values of $p$. The magnetic field and the
magnetization per site are plotted in units of $J/g\mu_{\rm B}$ and
$g\mu_{\rm B}$, respectively.}
\label{mag} 
\end{figure}

\subsection{Magnetization Curve}

 The magnetization curves at $T=0$ is calculated by the finite size DMRG
method for $2N=100$ for $\JA=2$ and 4 with $\JF=-2$ as shown in Fig.
\ref{mag}(a) and  (b), respectively. The average is taken over 100 samples.
A big plateau appears at $M = M_s(1-p)$ where $M_s$ is the saturated
magnetization and small jumps appear below and above the main plateau. 
Again, the physical interpretation of this structure can be given in terms
of the $q$-clusters. 
On the main plateau the spins connected by $\JA$-bonds form singlet clusters
and remaining spins are polarized along the direction of the magnetic field.
These plateaux are well quantized so that it must be useful to determine $p$
directly from experimental data. Similar fractional plateaux has been also
found in random polymerized XXZ chains\cite{cabra}.

Recently, Totsuka\cite{totsuka} analytically discussed the effect of
randomness on the magnetization plateaux. At first sight, our fractional
plateau appears to contradict with his criterion (Eq. (36) of ref.
\citen{totsuka}). However,  his argument concerns the fate of the plateaux
which already exist in the regular system and does not rule out the plateaux
induced by the bond randomness in our model as well as in the model in ref.
\citen{cabra} as commented by Totsuka himself at the end of his
paper\cite{totsuka}.

\begin{figure}
\centerline{\includegraphics[width=70mm]{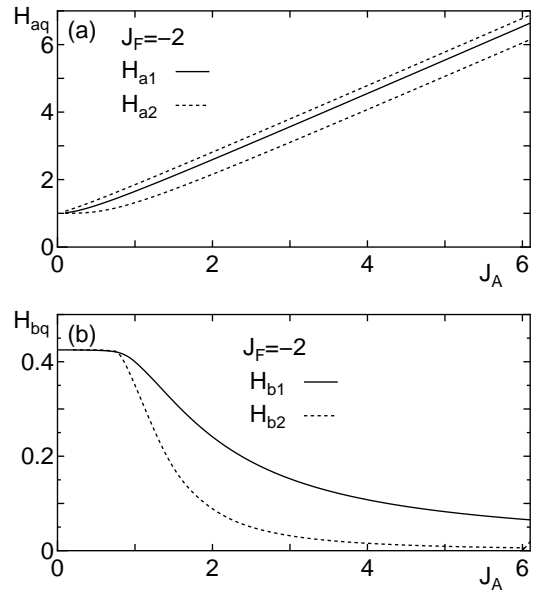}}
\caption{The $\JA$-dependence of the magnetic fields where the jumps due to
$q=1, 2$ clusters take place (a) above the plateau (solid line: $H_{\rm a1}$,
two dotted lines:  $H_{\rm a2}$) and (b) below the plateau (solid line:
$H_{\rm b1}$, dotted line: $H_{\rm b2}$). $\JF=-2$.}
\label{jumpvar}
\end{figure}

The small jumps above the plateau are the contribution from the spins
connected by the $\JA$ bond and those below the plateau are the contribution
from the pairs of almost free spins which support the low energy magnetic
exciations.  The position of these jumps are identified by the DMRG
calculation for the $q$-clusters as indicated by the up and down directed
triangles in Fig. \ref{mag}(a) and (b) for each $q$. The position of the
main jumps above the plateau  $H_{\rm aq}$ and those below the plateau
$H_{\rm bq}$ due to the $q$-cluster are depicted in  Fig. \ref{jumpvar}
against $\JA$ for $q=1$ and 2. Because $H_{\rm a1}$ is sensitive to $\JA$, we could
determine  $\JA$ from  experimental value of  $H_{\rm a1}$. However,
considering that the saturation field of the $x=1$ compound ($p=0$) is
already above 40T\cite{manakasat}, the observation of this jump would be
rather difficult within the presently available pulse magnetic field $\sim$
80T unless $\JA$ is relatively small. Although less precise, $\JA$ could
 be also determined from experimental data for $H_{\rm b1}$ which should be
observable within the presently available magnetic field. In recent
magnetization measurement\cite{manakanew}, however, no prominant structure
is observed in real IPACu(Cl$_x$Br$_{1-x}$) for $x \geq 0.87$  in the low
field regime. The origin of this discrepancy is unclear at present.

\section{Effect of Three Dimensionality - Disorder Induced Magnetic Order}

As demonstrated above, the low energy sectors of the present model is
dominated by the almost free spins in $q$-clusters with large $q$. In the
presence of the interchain coupling, we therefore expect that the 3
dimensional network of these spins sustain the magnetic order observed in
the experiment mediated by the interaction with spins in  neighbouring
chains. Let us assume the  Hamiltonian with the interchain coupling as follows,
\begin{eqnarray}
H &=& \sum_{j} [\sum_{i=1}^{N}J  \v{S}_{2i-1,j} \v{S}_{2i,j} +
\sum_{i=1}^{N}J_i  \v{S}_{2i,j} \v{S}_{2i+1,j}] \nonumber \\ 
&+&\sum_{i=1}^{2N}\sum_{<j,j'>}J_{\perp}  \v{S}_{i,j} \v{S}_{i,j'}, 
\end{eqnarray}
where $j$ and $j'$ distinguish the chains and $\sum_{<j,j'>}$ denotes the
summation over nearest neighbour chains. We assume that the correlation
between two almost free spins separated by the $\JA$ bonds is mainly
mediated by the F-AF part of the neighbouring chains, because the
correlation length of the AF-AF part is small compared to that of the F-AF
part. Furthermore, for small $p$, the probability to find the $\JA$ bonds on
the neighbouring chains  between two almost free spins  is small. As a
result, the interchain interaction and the short range correlation within
the neighbouring chains support the $\uparrow \uparrow \downarrow
\downarrow$-type long range order as far as $p$ is small.

\begin{figure}
\centerline{\includegraphics[width=70mm]{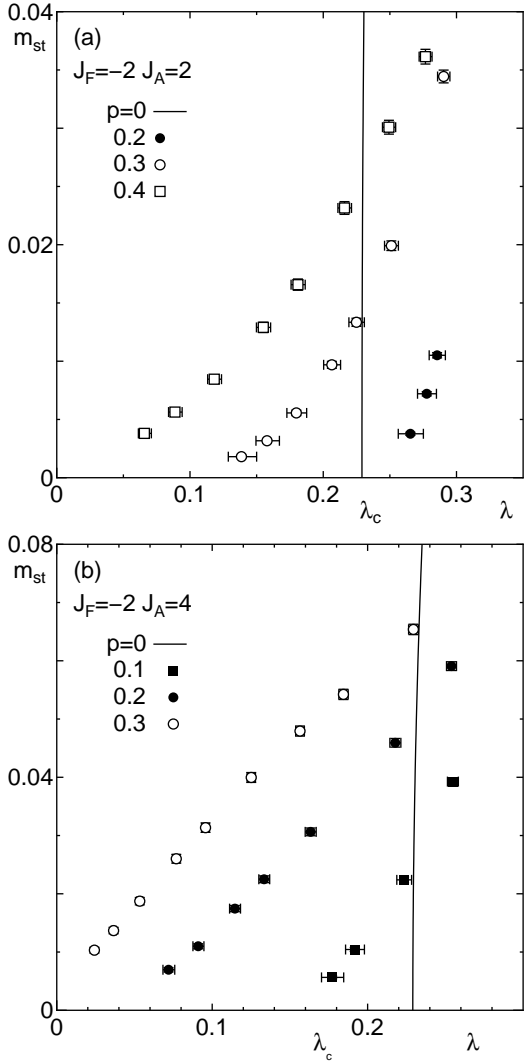}}
\caption{The $\lambda$-dependence of staggered magnetization for (a)$\JA=2$
and (b)$\JA=4$ with $\JF=-2$. The solid line is the staggered magnetization
for $p=0$.}
\label{magst}
\end{figure}

For quasi-one-dimensional systems, the mean field approximation for the
interchain coupling gives reliable results\cite{sca}, because  a large
number of spins are envolved in the low energy long wave length fluctuations
in each chain. These fluctuations are frozen by interchain interaction
resulting in the three dimensional ordering.  We therefore employ the
interchain mean field approximation assuming the  $\uparrow \uparrow
\downarrow \downarrow$ order to obtain the single chain mean field
Hamiltonian given by,
\begin{equation}
H^{MF} = \sum_{i=1}^{N}J  \v{S}_{2i-1} \v{S}_{2i} + \sum_{i=1}^{N}J_i
\v{S}_{2i} \v{S}_{2i+1} - \sum_{i=1}^{2N}h_i S_i^z
\end{equation}
with $h_{i}=-z_{\c}J_{\perp}<S_i^z>$ and $z_{\c}$ is the number of nearest
neighbour chains.  Reflecting the $\uparrow \uparrow \downarrow \downarrow$
order, we take $h_{4i}=h_{4i+1}=h$ and $h_{4i+2}=h_{4i+3}=-h$. In Fig.
\ref{magst}, the staggered magnetization $\mst \equiv
<\frac{1}{2N}\sum_{i=1}^{N}(-1)^i( S_{2i}^z + S_{2i+1}^z )>$ is plotted
against $\lambda \equiv z_{\c}J_{\perp}$ for (a) $\JA=2$ and (b) $\JA=4$
with $\JF=-2$. The chain length is $2N=200$ and  finite size DMRG method is
used. The average is taken over 200 samples.  For $p=0$, $\mst$ vanishes
unless $\lambda$ is larger than a critical value $\lambda_c$.  For finite
$p$, however, $\mst$ takes a finite value  even for small $\lambda <
\lambda_c$. The magnitude of $\mst$ increases with $p$ and $\JA$, namely
with the increase of randomness. It should be noted that the long range
order for $\lambda < \lambda_c$ starts to appear around $p \sim 0.3$ for
$\JA=2$ and  around $p \sim 0.1$ for $\JA=4$. Theses values are
approximately consistent with the estimation of $p_c$ from the energy gap
scaling. Therefore the $\uparrow \uparrow \downarrow \downarrow$-type long
range order is stabilized for appropriate strength of the interchain
coupling if the low energy density of state has divergent singularity as
expected.

 In the experiment, the long range order is observed for $x < 0.87$.  If we
assume $p=1-x^2$,  the regime $x <0.87$ corresponds to $p \simgeq 0.24$.
Although the exact value of $\JA$ is unknown,  we find from Fig \ref{magst}
that the long range order is stabilized in this concentration range even for
$\lambda < \lambda_{c}$ both for $\JA=2$ and 4.  These results explain the
qualitative features of the  experimental observation of the magnetic
ordered state. For the quantitative comparison with experiments, the
strength of the \clbr bonds need to be determined. Similar analysis is made
for the random $S=1$ Heisenberg antiferromagnet by Villar {\it et
al}\cite{vilar}.

\section{Summary and Discussion}

The $S=1/2$ Heisenberg chains  with bond alternation and randomness on the
strong bonds are studied by the DMRG method. The low energy spectrum is
shown to have the Griffiths type singularity and the physical origin of this
behavior is explained based on the $q$-cluster picture. In the magnetization
curves, there appears a randomness induced plateau at the fractional value
of the saturated magnetization. This plateau and the fine sructures of the
magnetization curves are also understood based on the $q$-cluster picture.
By the mean field treatment of the interchain coupling, the  magnetic long
range order is shown to be stabilized by randomness in the intermediate
concentration regime. The results are discussed in relation with the recent
experiment for (CH$_3)_2$CHNH$_3$Cu(Cl$_x$Br$_{1-x})_3$.

We carried out the interchain mean field calculation assuming the
$\uparrow\uparrow\downarrow\downarrow$ type long range order. However, the
possibility of different type of ordering such as spin glass ordering cannot
be excluded. From this viewpoint, the experimental determination of the
magnetic structure by neutron scattering experiment is also hoped to
elucidate the nature of the long range order. 

Recently, Nakamura\cite{naka} has shown that the
$\uparrow\downarrow\uparrow\downarrow$ correlation becomes most critical in the
intermediate concentration regime in the present model from the
nonequilibrium relaxation method analysis of the quantum Monte Carlo data.
We have also checked the possibility of the
$\uparrow\downarrow\uparrow\downarrow$ order within the DMRG and interchain
mean field approximation. According to our calculation, the staggered magnetization for $\uparrow\downarrow\uparrow\downarrow$ order is much lower than that for the $\uparrow\uparrow\downarrow\downarrow$ order. For example, for $\JA=-\JF=2$ and $p=0.4$, the staggered magnetization for the $\uparrow\downarrow\uparrow\downarrow$ order is approximately one order of magnitude smaller than that for the $\uparrow\uparrow\downarrow\downarrow$ order. For other values of parameters $\JA, \JF$ and $p$, the ratio of $\uparrow\downarrow\uparrow\downarrow$ order to $\uparrow\uparrow\downarrow\downarrow$ order is even smaller. Especially, the  $\uparrow\downarrow\uparrow\downarrow$ order decreases rapidly with the decrease of $p$ and becomes less than $10^{-3}$ for $p \leq 0.3$ with $\JA=-\JF=2$ and for  $p \leq 0.2$ with $\JA=4$, $\JF=-2$ within the regime $\lambda < \lambda_c$. Therefore we may conclude that the dominant order is of the $\uparrow\uparrow\downarrow\downarrow$ type although it is possible that the weak $\uparrow\downarrow\uparrow\downarrow$ order also coexist.
It should be
noted that Nakamura's calculation does not explicitly include the effect of
interchain coupling. In the random systems, the criticality of the
correlation in a single chain does not always imply the corresponding long
range order in its quasi-1-dimensional counterpart. 

The absence of the fine structure in the low field magnetization curve of
the real IPACu(Cl$_x$Br$_{1-x}$) is not understood in the present
calculation. In the presence of interchain coupling, however, it is possible
that the  almost free spins form localized singlet clusters mediated by
interchain coupling. This can suppress the low field structure in the
magnetization curve. However, such local correlation effect is not properly
described within the present interchain mean field approximation. This
problem is left for future studies.

In this work, we concentrated on the ground state properties  due to the
limitation of the numerical method (DMRG). The finite temperature effects
must be important for the direct comparison with experiments. This is left
for future studies. 

The author is grateful to H. Manaka for valuable discussion and for
explaining the details of the analysis of the experimental data. He also
thanks  T. Nakamura for useful discussion. The computation in this work has
been done using the facilities of the Supercomputer Center, Institute for
Solid State Physics, University of Tokyo and the Information Processing
Center, Saitama University.  This work is supported by a Grant-in-Aid for
Scientific Research from the Ministry of Education, Culture, Sports, Science
and Technology, Japan.

\end{document}